\documentclass[runningheads]{llncs}

\usepackage[T1]{fontenc}
\usepackage{graphicx}
\graphicspath{{../}}
\usepackage{booktabs}
\usepackage{amsmath}
\usepackage{amssymb}
\usepackage{multirow}
\usepackage{xcolor}
\usepackage{hyperref}
\usepackage{float}
\usepackage{subfigure} 
\usepackage{subcaption} 
\usepackage{cite}

\DeclareCaptionLabelSeparator{bar}{\space\textbf{|}\space}
\newcommand{\seniorauthor}{\textsuperscript{*}}
\begin{document}

\title{Deep learning-guided discovery reveals the diversity and mutational potential of environmental \texorpdfstring{$\beta$-lactamases}{beta-lactamases}}

\author{
Liangjun Ge
\and
Qiandi Gao
\and
Haiming Xu
\and
Zhiwei Qin\seniorauthor
\and
Heqian Zhang\seniorauthor
}

\authorrunning{Ge et al.}
\titlerunning{Ge et al.}

\institute{
Center for Biological Science and Technology, Advanced Institute of Natural Sciences, Beijing Normal University, Zhuhai, Guangdong, 519087, China  \\
\seniorauthor Correspondence: \texttt{z.qin@bnu.edu.cn}; \texttt{zhangheqian@bnu.edu.cn}
}

\maketitle

\begin{abstract}
$\beta$-lactamases are a major mechanism of bacterial resistance to $\beta$-lactam
antibiotics; however, their functional diversity and mutational
potential in environmental microbial communities remain partially
understood. This study aimed to characterize environmental $\beta$-lactamase
diversity and investigate how sequence variation influences antibiotic
resistance. By integrating protein language model (PLMs) with ensemble
learning, we identified more than 46,000 putative $\beta$-lactamases across
all four Ambler classes from environmental metagenomes, widely
distributed across diverse habitats, with distinct habitat-associated
class compositions. The candidates sharing approximately 30\% sequence
identity with known $\beta$-lactamases conferred resistance to multiple
$\beta$-lactam antibiotics when expressed in \emph{Escherichia coli}.
Combining inverse folding model with PLMs further identified mutations
that altered resistance profiles, including substitutions that increased
resistance to carbapenems. Predicted high-fitness sites exhibited
greater natural sequence variability overall, with higher variability in
wastewater-derived than ocean-derived enzymes. These findings reveal
undercharacterized $\beta$-lactamase diversity and substantial mutational
flexibility in environmental microbiomes, providing a basis for
investigating their potential contribution to the diversification of
antibiotic resistance.
    
\end{abstract}

\section{Introduction}

$\beta$-lactam antibiotics represent one of the most widely used classes of
antimicrobial agents in clinical practice~\cite{Bush2016BetaLactams}. Their
primary mechanism of action involves the inhibition of
penicillin-binding proteins (PBPs), particularly their transpeptidase
activities responsible for cross-linking peptidoglycan strands during
bacterial cell wall biosynthesis. By structurally mimicking the terminal
D-Ala-D-Ala motif of peptidoglycan precursors, $\beta$-lactam antibiotics bind
to and covalently inhibit PBPs, thereby compromising bacterial cell wall
integrity and ultimately leading to bacterial cell death~\cite{MoraOchomogo2021Targets}(Mora-Ochomogo
and Lohans 2021). However, the clinical effectiveness of $\beta$-lactam
antibiotics is increasingly threatened by the emergence and widespread
dissemination of $\beta$-lactamases, enzymes capable of hydrolyzing the
$\beta$-lactam ring and inactivating these antibiotics~\cite{Tooke2019BetaLactamases,Bush2020Epidemiology}.

To date, $\beta$-lactamases have been classified into four major classes (A,
B, C, and D) according to the Ambler system, primarily based on amino
acid sequence similarity. Classes A, C, and D comprise serine
$\beta$-lactamases (SBLs), which employ a conserved Ser-X-X-Lys motif in their
catalytic process, while Class B enzymes are metallo-$\beta$-lactamases (MBLs)
that rely on zinc ions for hydrolysis~\cite{Tooke2019BetaLactamases,Ambler1980Structure}. The
complementary Bush-Jacoby system further classifies $\beta$-lactamases
according to their substrate hydrolysis profiles and inhibitor
susceptibility~\cite{Bush2011Expansion}. Despite substantial progress in
$\beta$-lactamase classification and characterization, the diversity of these
enzymes continues to expand as additional sequences are discovered in
clinical and environmental samples~\cite{Gatica2019Comparative,Gholipour2026Resistome}. In particular,
the functional diversity of $\beta$-lactamases in natural microbial
communities remains incompletely characterized.

Environmental metagenomic studies have revealed that $\beta$-lactamase
homologs are widely distributed across diverse habitats, including
soils~\cite{Gudeta2016Metagenomics,Allen2009Soil}, aquatic ecosystems~\cite{Cui2022Aquaculture},
wastewater~\cite{Munk2022Sewage,Makowska2020Wastewater,Knight2025Continuum}, and host-associated microbial
communities~\cite{Economou2015Agriculture,Jia2024Microbiome}. Previous studies have identified
functionally active $\beta$-lactamases with substantial sequence divergence
from previously characterized enzymes~\cite{Sommer2009Reservoir}. Comparative
metagenomic analyses have also shown differences in the diversity and
distribution of $\beta$-lactamase homologs across environmental habitats,
including marine and wastewater-associated ecosystems~\cite{Gatica2019Comparative}. Recently, divergent $\beta$-lactamases have been identified in the
human microbiome and experimentally characterized~\cite{Jia2024Microbiome}.
Nevertheless, the systematic identification and functional
characterization of $\beta$-lactamases from large-scale environmental
metagenomes remain challenging, particularly for divergent sequences
with low similarity to known enzymes. Moreover, the ecological
distribution, functional properties, and mutational potential of these
environmental $\beta$-lactamases remain insufficiently understood. Combining
computational prediction with experimental validation may provide a more
comprehensive understanding of their diversity and functional
variability.

In this study, we aim to systematically investigate the diversity,
ecological distribution, functional properties, and evolutionary
characteristics of $\beta$-lactamases in environmental microbial communities.
Specifically, our objectives are: (i) to train a deep learning-based
model integrating protein language models (PLMs) and ensemble learning
to identify and classify $\beta$-lactamase candidates from large-scale
environmental metagenomic datasets; (ii) to investigate their ecological
distribution and habitat-associated diversity, and experimentally
characterize representative divergent $\beta$-lactamases; (iii) to combine
protein inverse folding and PLMs to predict potential high-fitness
mutations, followed by experimental validation and molecular dynamics
simulations to explore their functional effects and possible structural
mechanisms; and (iv) to compare sequence variation at predicted
high-fitness sites among $\beta$-lactamases from different environmental
habitats and explore its potential association with environmental
selective pressures. Our work expands our understanding of environmental
$\beta$-lactamase diversity, functional properties, and sequence variation,
and provides a basis for further investigating environmental
resistance-associated genes and their potential risks.

\section{Results}
\subsection{Phylogenetic Diversification and Clade-Specific Functional Phenotypes of Natural $\beta$-Lactamases}
\vspace{-2mm}

\begin{figure}[!htbp]
  \centering
  \includegraphics[width=\linewidth]{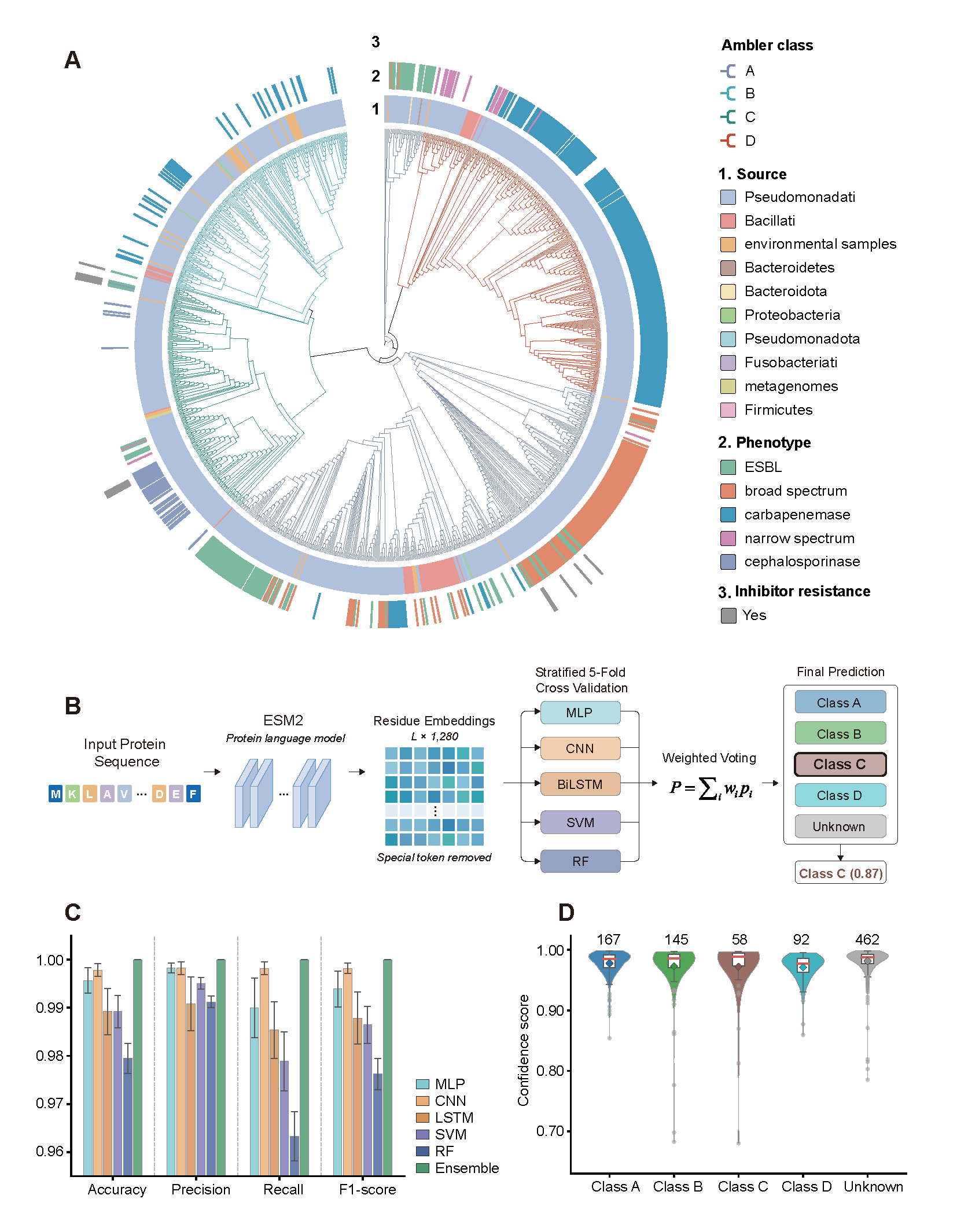}
    \setlength{\abovecaptionskip}{1pt}
    \caption{\scriptsize \textbf{Phylogenetic diversity of known $\beta$-lactamases and the
ensemble-learning framework.} \textbf{(A)} Maximum-likelihood phylogeny
of 1,066 naturally occurring $\beta$-lactamases, annotated with Ambler class,
source, and functional phenotype. \textbf{(B)} Ensemble-learning
framework integrating ESM2-derived protein representations with five
classifiers. \textbf{(C)} Classification performance of the individual
classifiers and ensemble model across 5-fold cross-validation.
\textbf{(D)} Distribution of the ensemble model's confidence scores for
correctly classified samples across the complete dataset.}
  \label{fig:phylogenetic_tree}
\end{figure}

We assembled a dataset of 1,778 experimentally characterized $\beta$-lactamase
sequences from the BLDB and Swiss-Prot databases~\cite{Naas2017BLDB,UniProt2025Knowledgebase}. By Ambler class, the sequences comprised class A (n = 803), class B
(n = 319), class C (n = 275), and class D (n = 381); 1,066 were of
natural origin, and 712 were engineered. To investigate the sequence
diversification of natural $\beta$-lactamases and its relationship with host
taxonomy and functional phenotype, we inferred a maximum-likelihood
phylogenetic tree from the 1,066 naturally occurring sequences,
supplemented with annotation information from BLDB and
NCBI~\cite{OLeary2024Datasets} (\textbf{Fig.~\ref{fig:phylogenetic_tree}A}). The phylogeny showed that
the four Ambler classes formed well-separated clades, indicating
pronounced sequence divergence among classes. This pattern is consistent
with the well-established distinction between serine $\beta$-lactamases
(classes A, C, and D) and metallo-$\beta$-lactamases (class B) in catalytic
residue composition and mechanism. A partial correspondence was observed
between phylogenetic clades and microbial taxonomic groups; however,
instances of the same species spanning multiple clades were also
detected. These observations suggest that, in addition to vertical
inheritance, horizontal gene transfer, gene duplication, and the
coexistence of distinct enzyme subtypes within a single host may all
contribute to the dissemination and diversification of $\beta$-lactamase
genes.

The distribution of functional phenotypes displayed clade-level
specificity. The carbapenemase phenotype was concentrated in classes B
and D, the extended-spectrum $\beta$-lactamase (ESBL) phenotype occurred
predominantly within class A clades, and the inhibitor-resistance
phenotype was restricted to classes A and C. This distribution
underscores the need for effective inhibitors targeting class B and
class D $\beta$-lactamases. Furthermore, broad-spectrum, ESBL,
carbapenemase, and inhibitor-resistance phenotypes co-occurred within
class A clades, suggesting that class A $\beta$-lactamases may harbour more
complex and clinically noteworthy combinations of resistance
phenotypes. Taken together, these findings indicate that different
$\beta$-lactamase lineages may possess distinct functional evolutionary
potentials and exhibit considerable functional plasticity under
antibiotic and inhibitor selective pressures. Notably, sequences
derived from environmental samples and metagenomic sources were
distributed across multiple clades in the phylogeny, implying that
diverse and widespread $\beta$-lactamase gene pools exist in natural
environments and that environmental microbiota may constitute an
important reservoir of novel $\beta$-lactamases.

\vspace{-3mm}

\vspace{-3mm}
\subsection{An Ensemble-Learning Framework for Potential $\beta$-Lactamases Identification}
\vspace{-2mm}

Protein language models (PLMs) learn and encode features related to
protein function and evolution from large-scale sequence data,
facilitating the identification of distantly related homologues, and
have been widely adopted for protein representation and function
prediction~\cite{Huang2026DeepAden,Gao2025FewShot}. Inspired by this, we developed an
ensemble learning framework based on a PLM to identify potential novel
$\beta$-lactamases from environmental metagenomic data (\textbf{Fig.~\ref{fig:phylogenetic_tree}B}).
ESM2-t33-650M was employed as the feature extraction
backbone~\cite{Lin2023Structure}. For each input protein sequence, the ESM2
model first generated a 1,280-dimensional embedding for every amino acid
residue. Special tokens at the sequence termini were then removed to
prevent them from confounding downstream classification, retaining only
residue-level representations for subsequent analysis.

To exploit the ESM2-derived feature sequences, we designed five
classifiers with complementary modelling properties to discriminate
among Ambler classes of $\beta$-lactamases and negative samples. Three were
deep learning classifiers: a multilayer perceptron (MLP), a
convolutional neural network (CNN), and a bidirectional long short-term
memory network (BiLSTM). The remaining two were conventional machine
learning classifiers: a support vector machine (SVM) and a random forest
(RF). The MLP was intended to capture global sequence
features~\cite{Son2024BASE}; the CNN integrated multi-scale information
through feature concatenation and max-pooling operations, which is
expected to recognize conserved motifs and local structural
signatures~\cite{Li2020MotifCNN}; and the BiLSTM used bidirectional sequence
processing to capture contextual dependencies across the
sequence~\cite{Shi2021DeepLearning}. By contrast, SVM and RF generally exhibit
greater robustness and generalisation under small-sample conditions~\cite{Xu2021DeepNetworks}. To combine the complementary strengths of these
classifiers, a soft-voting strategy was adopted: the class prediction
probabilities from each classifier were averaged with assigned weights,
and the class with the highest aggregate probability was selected as the
final prediction (\textbf{Fig.~\ref{fig:phylogenetic_tree}B}).

Each classifier was trained and evaluated using 5-fold cross-validation.
Stratified sampling was applied during data partitioning to ensure that
the class distribution in every fold matched that of the original
dataset, thereby reducing evaluation bias caused by class imbalance. All
five base classifiers achieved good classification performance, although
their predictions varied across validation folds (\textbf{Fig.~\ref{fig:phylogenetic_tree}C}).
The ensemble model, by contrast, correctly classified every validation
sample across all folds and delivered predictions with high confidence
(\textbf{Fig.~\ref{fig:phylogenetic_tree}C and D}). These results indicate that the proposed
ensemble framework is stable and reliable, capable of accurately
identifying members of the $\beta$-lactamase superfamily and predicting their
Ambler class, thus providing an effective tool for mining potential
$\beta$-lactamases from environmental metagenomic data.

\vspace{-3mm}
\subsection{Global Mining and Ecological Distribution of Putative $\beta$-Lactamases in MGnify}
\vspace{-2mm}

\begin{figure}[!t]
  \centering
  \includegraphics[width=\linewidth]{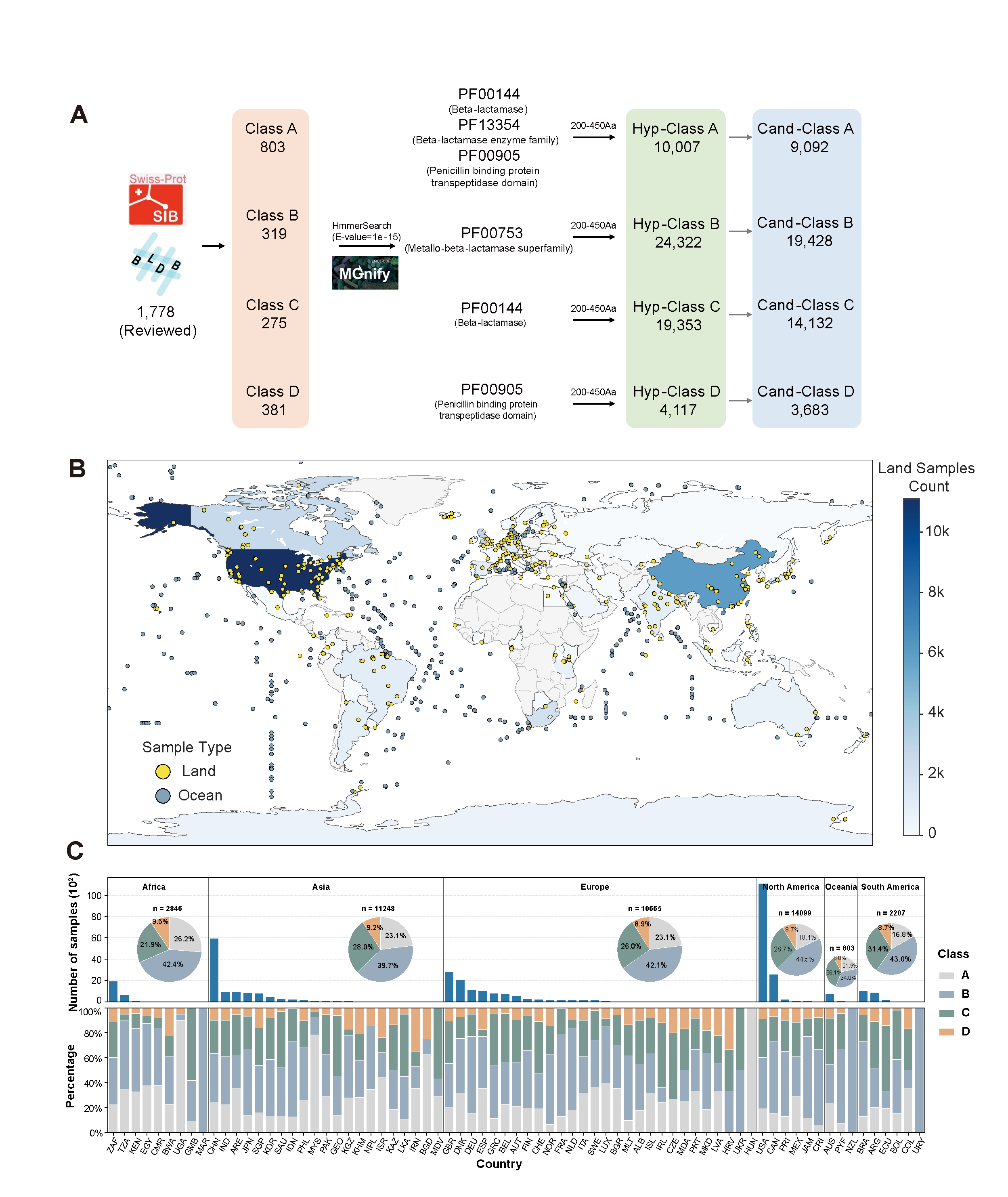}
    \setlength{\abovecaptionskip}{1pt}
    \caption{\scriptsize \textbf{Mining and global distribution of putative
environmental $\beta$-lactamases.} (A) Numbers and Ambler class composition of
candidate sequences identified from MGnify through HMM screening and
ensemble-model filtering. \textbf{(B)} Global distribution of candidates
with available geographic coordinates. Yellow and blue circles indicate
land and ocean sampling locations, respectively. Country shading
represents the number of land samples, with darker blue indicating
higher sample counts. \textbf{(C)} Numbers and Ambler class composition
of candidate $\beta$-lactamases across continents and countries.}
  \label{fig:global_distribution}
\end{figure}

The ensemble learning framework was next applied to mine potential
$\beta$-lactamases from the MGnify Protein Database (release
2024\_04)~\cite{Richardson2023MGnify}. Hidden Markov model (HMM) profiles were
first constructed for each Ambler class and used to scan the entire
database with HMMER (\textbf{Fig.~S1}). Sequences that passed this
initial screen were further filtered by length, retaining only those
within the 200--400 aa range observed across the 1,778 reference
sequences. All four Ambler classes yielded large numbers of hypothetical
sequences, and the retained hypotheses preserved the conserved domain
signatures typical of $\beta$-lactamases, with clear structural differences
among classes. The ensemble framework was then applied to these initial
hits to refine the predictions and reduce the false positive rate. After
ensemble filtering, each Ambler class retained a set of high-confidence
candidates, with classes B and C showing greater diversity and richness.
Notably, the class composition of these candidates did not match the
distribution of the known reference sequences, in which class A
predominates (\textbf{Fig.~\ref{fig:global_distribution}A}). This discrepancy suggests that the
observed class distribution of characterised $\beta$-lactamases may be
influenced by the focus of clinical research and may not fully represent
the true diversity of $\beta$-lactamases in natural environments.

Using MGnify samples with available geographic coordinates and habitat metadata, 
the predicted candidate sequences were examined for their global distribution and 
habitat specificity. Candidate sequences were widely distributed across terrestrial 
and marine environments worldwide, but their spatial distribution was markedly uneven. 
The majority of candidates originated from North America (n = 14,099), Asia (n = 11,248), 
and Europe (n = 10,665), with the United States contributing the largest number of samples; 
sequences from Africa, South America, and Oceania were comparatively sparse (\textbf{Fig.~\ref{fig:global_distribution}B}). 
This unevenness likely reflects both genuine differences in environmental microbial community composition 
across regions and biases in metagenomic sampling and sequencing effort. At the class level, 
the composition of candidates was broadly consistent across continents. 
Class B predominated in all regions (34.0-44.5\%), supporting the inference from the phylogenetic analysis 
that environmental microbiomes may serve as important reservoirs of class B $\beta$-lactamases, a pattern consistent 
with the observation that many known class B enzymes originate from environmental samples, 
as presented in \textbf{Fig.~\ref{fig:phylogenetic_tree}A}. Class C was the second most abundant category and showed relatively higher proportions 
in South America and Oceania, suggesting possible regional ecological preferences. 
Class A was consistently detected across all regions, whereas class D consistently 
accounted for the smallest proportion with little inter-regional variation, 
indicating a lower detection rate of class D enzymes in environmental metagenomes (\textbf{Fig.~\ref{fig:global_distribution}C}).

\begin{figure}[!htbp]
  \centering
  \includegraphics[width=\linewidth]{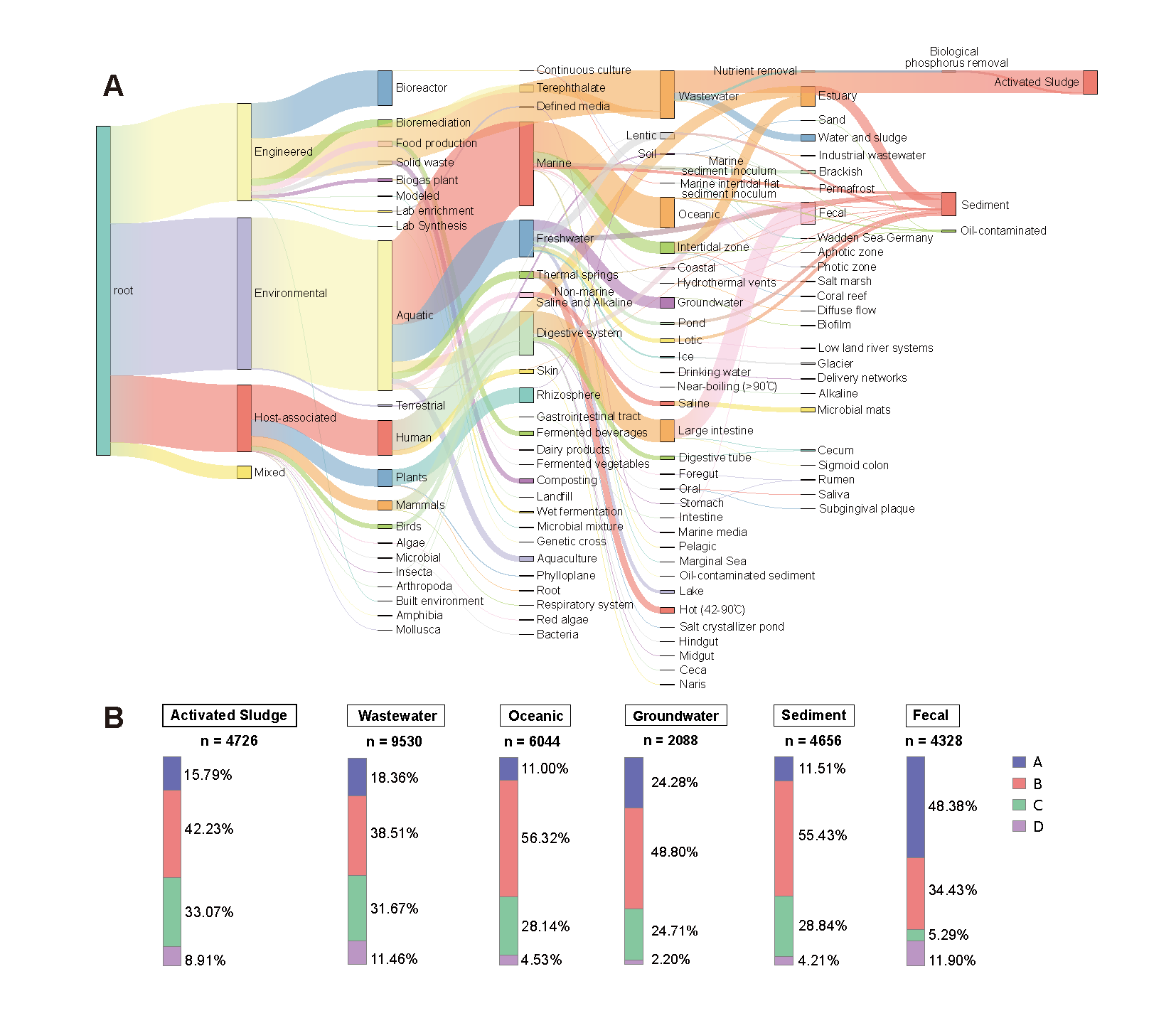}
    \setlength{\abovecaptionskip}{1pt}
    \caption{\scriptsize \textbf{Habitat distribution and class composition of putative
$\beta$-lactamases.} (A) Sankey diagram showing distribution of candidate
sequences across engineered water-treatment systems, natural aquatic
ecosystems, and host-associated microbiomes. \textbf{(B)} Relative
proportions of the four Ambler classes across some representative
habitats, illustrating habitat-associated differences in candidate
composition.}
  \label{fig:habitat_distribution}
\end{figure}

Further analysis of habitat-level distribution revealed that the candidate 
sequences were primarily associated with three broad habitat categories: 
engineered water-treatment systems, natural aquatic ecosystems, and host-associated microbiomes. 
Within engineered environments, candidate sequences were concentrated in activated sludge, 
which receives inputs from bioreactors, wastewater, and nutrient-removal treatment units. 
Natural aquatic environment candidates were predominantly derived from marine, freshwater, 
and sediment habitats. Host-associated candidates originated mainly from the human digestive system, 
particularly faecal and large-intestine samples (\textbf{Fig.~\ref{fig:habitat_distribution}A}). The class composition varied across 
habitats in ways that suggest ecological differentiation. Class B maintained a high proportion across 
diverse environments; combined with its high sequence diversity among environmentally sourced microbes, 
this pattern indicates that class B enzymes may possess broad environmental adaptability and 
wide ecological distribution (\textbf{Fig.~\ref{fig:habitat_distribution}B}). Both wastewater and activated sludge exhibited co-enrichment 
of class B and class C sequences (\textbf{Fig.~\ref{fig:habitat_distribution}B}). Wastewater continuously receives microbes and 
chemical substances from domestic, clinical, industrial, and environmental run-off, 
while activated sludge is characterised by high cell density, complex community interactions, 
and strong capacity for substance accumulation; these conditions may together favour 
the maintenance, enrichment, and potential horizontal gene transfer of multiple $\beta$-lactamase classes. 
By contrast, faecal samples displayed a distinct class composition dominated by class A (48.38\%) 
with a relatively high proportion of class D (11.9\%), suggesting that the host gut environment may 
selectively enrich class A and class D enzymes (\textbf{Fig.~\ref{fig:habitat_distribution}B}). The gut microbiota is shaped over time by 
host physiology, antibiotic exposure, and microbial competition, and its candidate sequence composition 
therefore differs markedly from that of open environments such as marine, sediment, and wastewater habitats. 
The pronounced enrichment of class A in faecal samples is consistent with the relatively high clinical attention 
given to class A enzymes in host-associated bacteria.

\vspace{-3mm}

\vspace{-3mm}
\subsection{Experimental validation confirms candidate $\beta$-lactamase activities}
\vspace{-2mm}

\begin{figure}[!htbp]
  \centering
  \includegraphics[width=\linewidth]{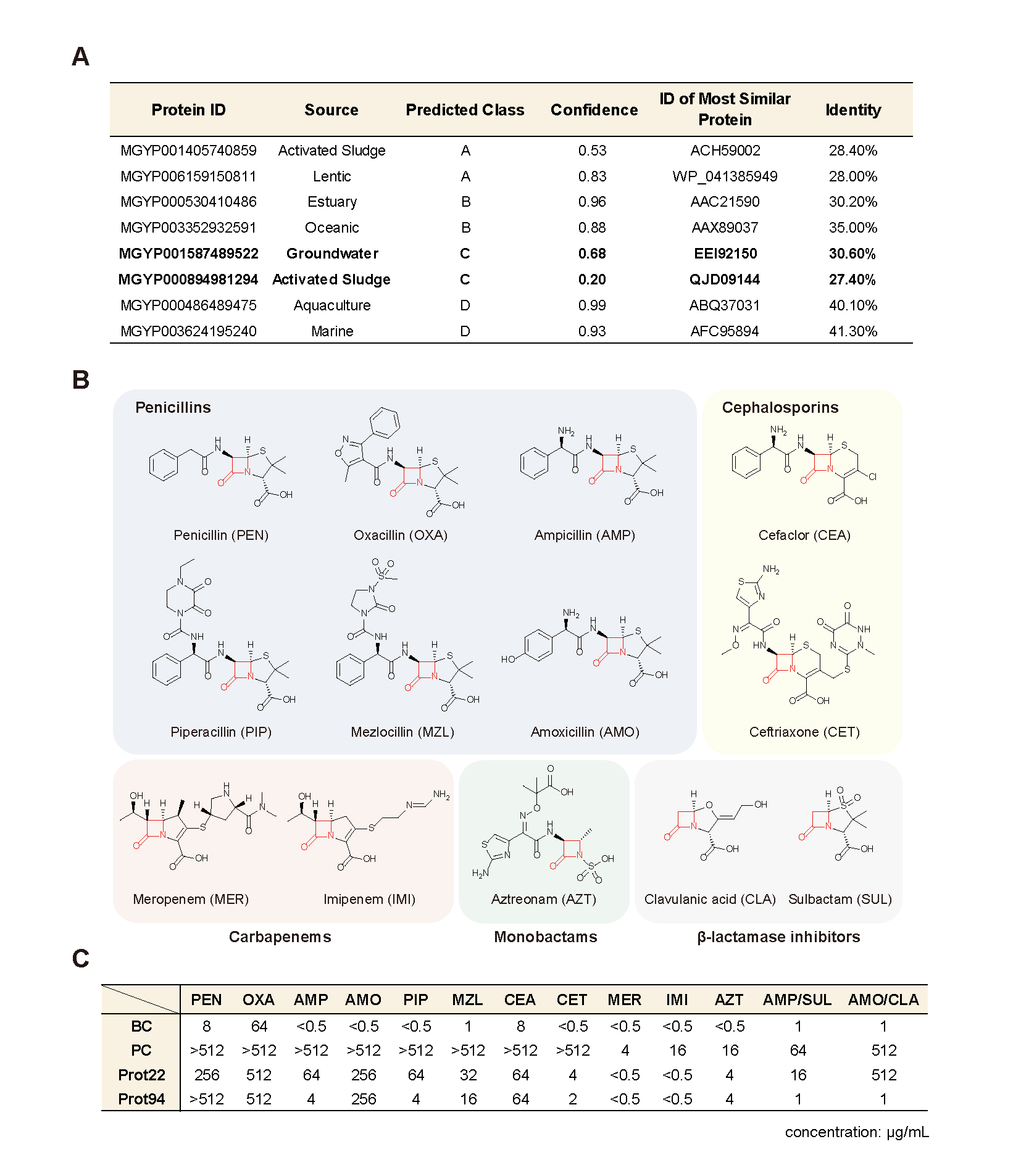}
    \setlength{\abovecaptionskip}{1pt}
    \caption{\scriptsize \textbf{Experimental validation of environmental $\beta$-lactamase
candidates.} (A) Two candidate $\beta$-lactamases from each Ambler class with
low sequence identity to known enzymes. Candidates selected for
experimental validation are shown in bold (Prot22 and Prot94).
\textbf{(B)} $\beta$-Lactam antibiotics and $\beta$-lactam/$\beta$-lactamase inhibitor
combinations tested. \textbf{(C)} Minimum inhibitory concentrations
(MICs) for \emph{E. coli} expressing Prot22, Prot94, or the positive
control AmpC, alongside the empty-vector control. PC, positive control;
BC, empty-vector control.}
  \label{fig:experimental_validation}
\end{figure}

To determine whether the candidate $\beta$-lactamases identified by the
ensemble learning framework exhibited functional activity, two
candidates with low similarity to the reference database were first
selected from each Ambler class. Among these, two class C candidates
derived from distinct antibiotic-pressure environments and exhibiting
low homology to known $\beta$-lactamases were chosen for experimental
validation: MGYP001587489522 (Prot22) and MGYP000894981294 (Prot94)
(\textbf{Fig.~\ref{fig:experimental_validation}A}). The characterised class C $\beta$-lactamase AmpC
(UniProt accession: P85302) served as a positive control (PC), and
\emph{E. coli} harbouring the empty pET26b vector served as a blank
control (BC)~\cite{Jia2024Microbiome}. To ensure uniform heterologous
expression conditions, the coding sequences of AmpC, Prot22, and Prot94
were codon-optimised; native signal peptides were removed during
cloning, and the optimised genes were inserted into the pET26b
expression vector, relying exclusively on the vector-encoded pelB signal
peptide for periplasmic targeting. Following transformation and
selection, AmpC, Prot22, and Prot94 were all stably expressed in
\emph{Escherichia coli} (\textbf{Fig.~S2}).

The resistance phenotypes of the recombinant strains were then assessed
by determining minimum inhibitory concentrations (MICs, $\mu$g/mL) against a
panel of $\beta$-lactam antibiotics. The tested antibiotics comprised
narrow-spectrum penicillins (penicillin and oxacillin), broad-spectrum
penicillins (ampicillin and amoxicillin), extended-spectrum penicillins
(piperacillin and mezlocillin), cephalosporins (cefaclor and
ceftriaxone), carbapenems (meropenem and imipenem), and a monobactam
(aztreonam). Two $\beta$-lactam/$\beta$-lactamase inhibitor combinations,
amoxicillin/clavulanic acid and ampicillin/sulbactam, were also tested
(\textbf{Fig.~\ref{fig:experimental_validation}B}). All agents were assayed across 11 two-fold
dilutions ranging from 0.5 to 512 $\mu$g/mL.

The blank control strain BC showed no appreciable resistance to any of 
the antibiotics or combination agents tested, exhibiting at most low-level 
tolerance. In contrast, Strains expressing AmpC, Prot22, or Prot94 each displayed 
$\beta$-lactam resistance phenotypes of varying magnitude. The positive control PC showed 
elevated MICs against most penicillins and cephalosporins but remained relatively 
susceptible to carbapenems and the monobactam. Prot22 exhibited marked penicillin 
resistance, with MICs exceeding 256 $\mu$g/mL for penicillin, oxacillin, ampicillin, 
and the amoxicillin/clavulanic acid. Prot94 displayed a resistance profile broadly 
similar to that of Prot22, with the highest MIC observed for penicillin (512 $\mu$g/mL), 
but showed greater susceptibility to the inhibitor combinations (\textbf{Fig.~\ref{fig:experimental_validation}C}). 
These results demonstrate that Prot22 and Prot94 each confer resistance to 
multiple $\beta$-lactam antibiotics when expressed in E. coli, confirming that the two low-homology 
class C candidates identified by the ensemble learning framework possess functional $\beta$-lactamase activity.

\vspace{-3mm}

\vspace{-3mm}
\subsection{AI-Guided Screening Strategy Identifies High-Fitness Mutations of $\beta$-Lactamase}
\vspace{-2mm}

Having confirmed the catalytic activity of the candidate $\beta$-lactamases,
we next used these two enzymes as models to explore potential
high-fitness (HF) mutation and to evaluate the utility of AI-assisted
protein engineering strategies for $\beta$-lactamase function optimization and
evolutionary potential assessment. Protein inverse-folding models, such
as ProteinMPNN, can efficiently sample sequence space under structural
constraints and have been shown to identify key functional mutations
that substantially enhance catalytic efficiency across diverse proteins,
including HF mutations in TEM-1 $\beta$-lactamase~\cite{Fei2025InverseFolding,Dauparas2022ProteinMPNN}.
Separately, protein language models from the ESM family learns
evolutionary constraints from natural protein sequences and enable
unsupervised prediction of the functional effects of amino acid
substitutions~\cite{Rives2021Scaling}. These models have been widely applied
to zero-shot mutation effect prediction~\cite{Brandes2023VariantEffects,Meier2021ZeroShot,Wang2025Maize}.

\begin{figure}[!htbp]
  \centering
  \includegraphics[width=\linewidth]{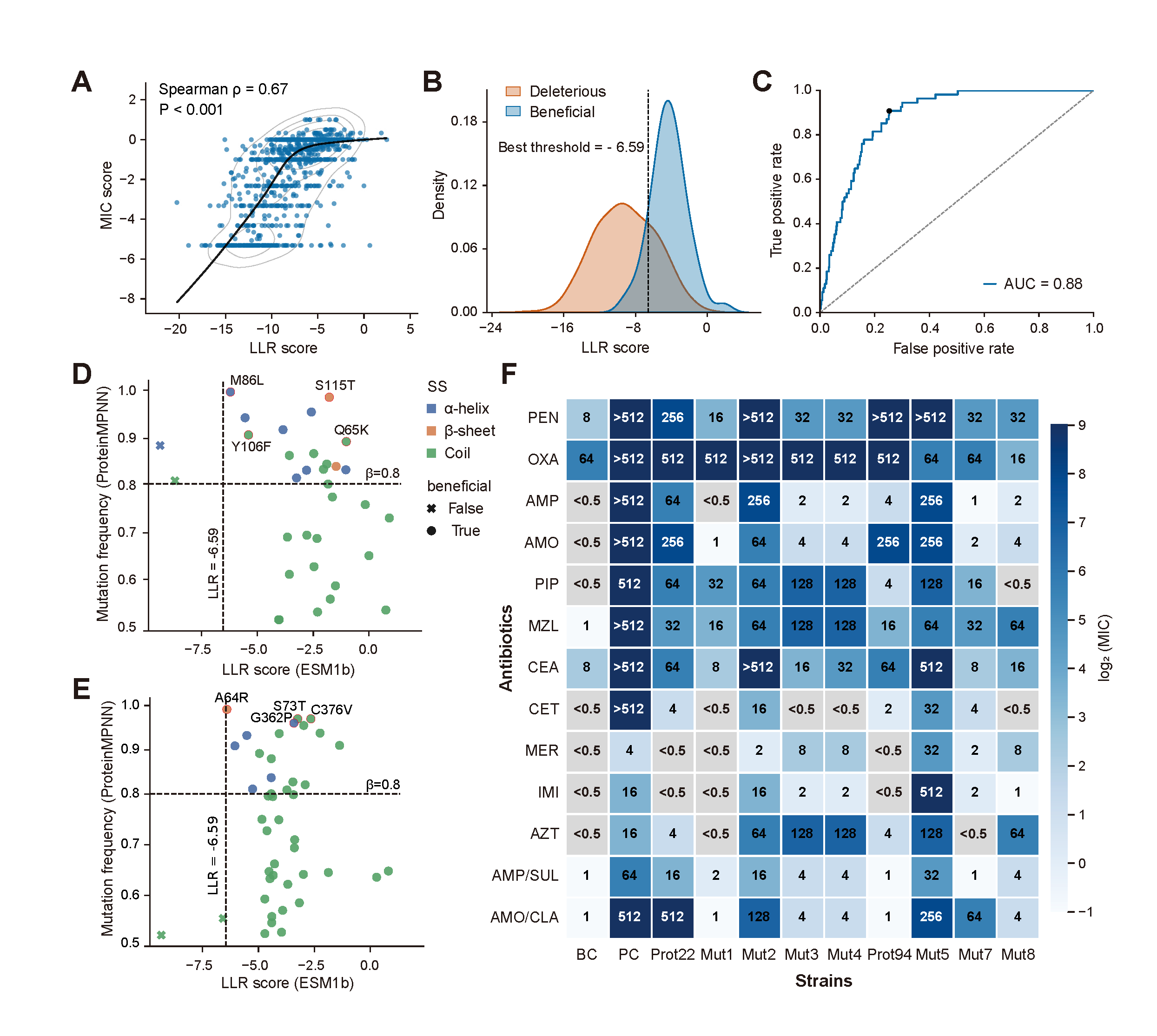}
    \setlength{\abovecaptionskip}{1pt}
    \caption{\scriptsize \textbf{AI-guided screening and experimental evaluation of
putative high-fitness mutations.} (A) Correlation between ESM1b-derived
log-likelihood ratio (LLR) scores and experimentally measured fitness of
TEM-1 mutations. \textbf{(B, C)} Discrimination of mutation effects by
ESM1b and selection of the LLR threshold using ROC analysis. \textbf{(D,
E)} Candidate mutations identified in Prot22 and Prot94 using
ProteinMPNN and ESM1b. Candidates selected for experimental validation
are marked with a red circle. \textbf{(F)} MIC profiles of the seven
successfully constructed mutants compared with their corresponding wild
types.}
  \label{fig:mutation_validation}
\end{figure}

To assess the applicability of ESM models to $\beta$-lactamase mutation effect prediction,
we computed the log-likelihood ratio (LLR) of each mutant residue relative to the wild-type
residue using ESM1b on a TEM-1 $\beta$-lactamase mutation dataset~\cite{Jacquier2013TEM1}, thereby quantifying the putative
fitness of each mutation. ESM1b-derived LLR scores showed a significant positive correlation with
experimentally measured mutation fitness (Spearman's $\rho$ = 0.667, \emph{p} < $2.97 \times 10^{-128}$; \textbf{Fig.~\ref{fig:mutation_validation}A}).
Beneficial mutations overall exhibited higher LLR scores, whereas deleterious mutations were
distributed in the lower-score range. ROC analysis further showed that ESM1b discriminated well
between beneficial and deleterious mutations (AUC = 0.88). Based on the maximum Youden index,
an LLR threshold of $-6.59$ was selected as the optimal threshold for identifying potentially
beneficial mutations (\textbf{Fig.~\ref{fig:mutation_validation}B and C}). These results demonstrate that ESM1b can provide reliable
evolutionary information for $\beta$-lactamase mutation prediction.

We therefore combined structure-constrained ProteinMPNN with
evolution-informed ESM1b to jointly identify HF mutation in Prot22 and
Prot94. Candidate mutations were evaluated according to
ProteinMPNN-derived mutation frequencies and ESM1b LLR scores. Following
the AiCE (AI-informed constraints for protein engineering) framework,
mutation acceptance thresholds were set at $\beta$ = 0.8 for rigid regions and
$\gamma$ = 0.5 for flexible regions~\cite{Fei2025InverseFolding}. Candidates with LLR
scores above $-6.59$ were retained, yielding 31 and 42 putative HF
mutation sites for Prot22 and Prot94, respectively (\textbf{Table~S1 and
S2}). Eight representative mutations located in rigid ($\alpha$-helix, $\beta$-sheet)
and flexible (Coil) regions were selected for experimental validation
(\textbf{Fig.~\ref{fig:mutation_validation}D and E}). Except for Mut6 (A64R), which failed sequencing
verification, all mutants were successfully constructed and subjected to
MIC testing.

All seven mutants displayed changes in MIC relative to the wild type,
including enhanced or reduced resistance and substrate spectrum shifts,
indicating that the predicted mutation sites can substantially affect
$\beta$-lactamase function. Among the Prot22-derived mutants, Mut2 (Y106F)
showed the most pronounced broad-spectrum resistance enhancement, with
increased MICs for penicillin, oxacillin, ampicillin, and cefaclor.
Notably, Mut2 also acquired carbapenem-related resistance phenotypes
absent from wild-type Prot22 (MER, 2 $\mu$g/mL; IMI, 16 $\mu$g/mL). Mut3 (M86L)
and Mut4 (Q65K) exhibited altered substrate profiles, with increased
MICs for piperacillin, mezlocillin, and aztreonam but reduced resistance
to certain penicillin substrates. In contrast, Mut1 (S115T) displayed an
overall reduction in resistance. Among the Prot94 mutants, Mut5 (S73T)
showed the strongest resistance enhancement, with MICs of 512 $\mu$g/mL for
both cefaclor and imipenem, and high-level resistance to most tested
$\beta$-lactam antibiotics. Conversely, Mut7 (C376V) and Mut8 (G362P)
exhibited reduced MICs for penicillins compared to wild-type Prot94.
(\textbf{Fig.~\ref{fig:mutation_validation}F}).

\vspace{-3mm}

\begin{figure}[!t]
  \centering
  \includegraphics[width=\linewidth,height=0.55\textheight,keepaspectratio]{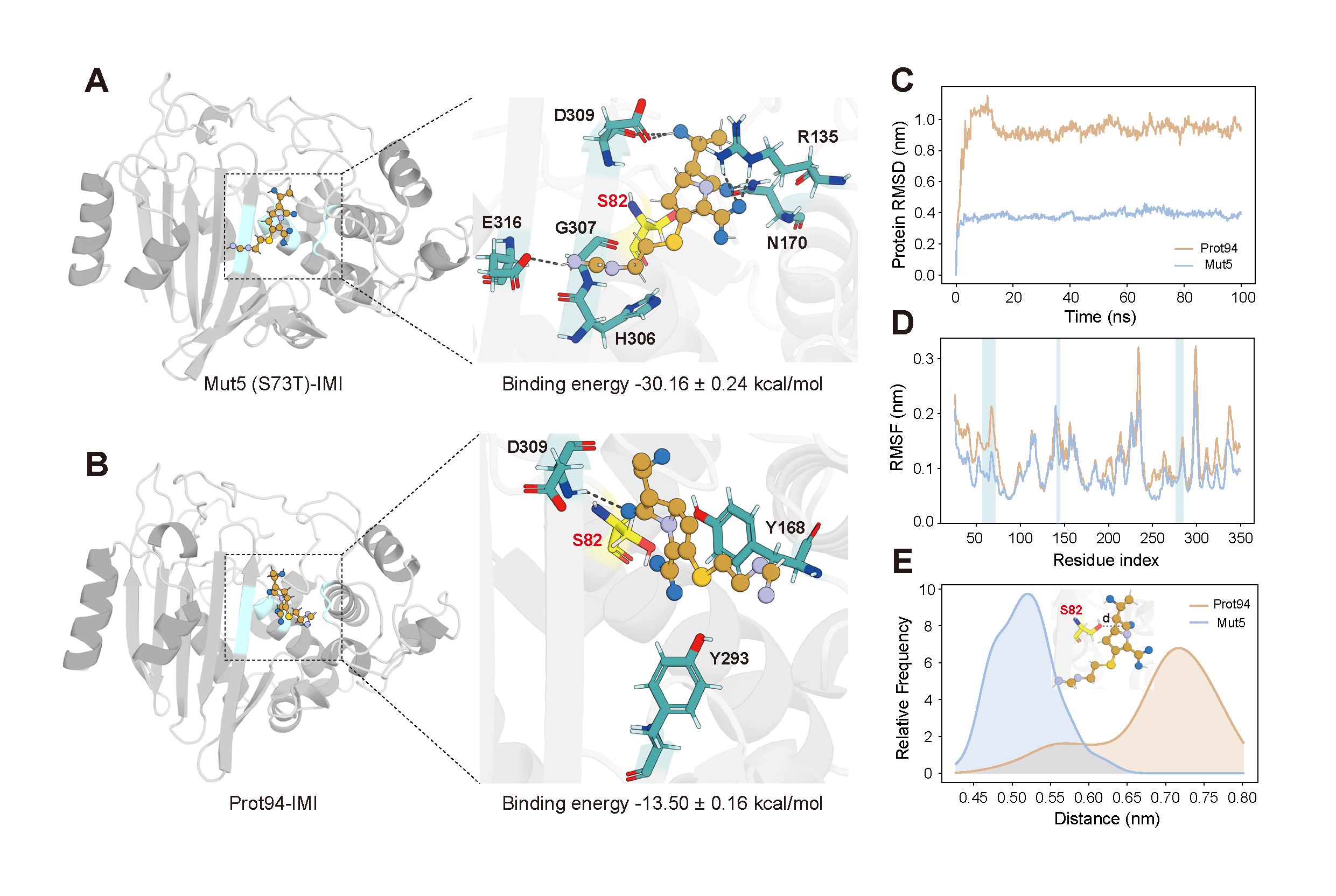}
    \setlength{\abovecaptionskip}{1pt}
    \caption{\scriptsize \textbf{Molecular dynamics (MD) analysis of Mut5 and wild-type
Prot94 enzyme-substrate complexes.} Comparison of intermolecular
hydrogen bonds between the mutant Mut5 \textbf{(A)} and the wild-type
Prot94 \textbf{(B)}. Substrates are shown as brown ball-and-stick
models, with hydrogen bonds indicated by grey dashed lines. Catalytic
serine and residues forming hydrogen bonds with the substrates are shown
as yellow and cyan sticks, respectively. \textbf{(C, D)}
Root-mean-square deviation (RMSD) and root-mean-square fluctuation
(RMSF) analyses during MD simulations. \textbf{(E)} Distribution of
distances (d) during MD simulations. The distance d represents the
separation between the $\beta$-lactam carbonyl carbon and the O$_\gamma$ atom of the
catalytic serine.}
  \label{fig:MD_analysis}
\end{figure}

To investigate the structural basis for the enhanced activity of Mut2 and Mut5,
100-ns molecular dynamics simulations were performed on the Mut2-CEA and Mut5-IMI
complexes and their corresponding wild-type complexes. Substrates were initially
docked into the active site, and the pose with the shortest distance between
the $\beta$-lactam carbonyl carbon and the O$_\gamma$ atom of the catalytic serine was selected
for simulation (\textbf{Fig.~\ref{fig:MD_analysis}A and B; Fig.~S3}). The Mut5-IMI complex exhibited lower RMSD values
and lower RMSF values for binding-pocket residues than the wild-type complex,
indicating that the S73T substitution improved both the global conformational stability
and local stability of the catalytic pocket (\textbf{Fig.~\ref{fig:MD_analysis}C and D}). In Mut5, IMI formed
a more extensive hydrogen-bond network with D309, R135, and E316, and the binding free energy
was lower than that of the wild type ($-30.16 \pm 0.24\,\mathrm{kcal/mol}$ compared with
$-13.50 \pm 0.16\,\mathrm{kcal/mol}$). In addition, the distance between the S82 O$_\gamma$
atom of the catalytic serine and the substrate carbonyl carbon was shorter in Mut5 than
in the wild type ($0.52 \pm 0.04\,\mathrm{nm}$ compared with $0.69 \pm 0.07\,\mathrm{nm}$),
suggesting a more favourable orientation for nucleophilic attack and acyl-enzyme intermediate
formation (\textbf{Fig.~\ref{fig:MD_analysis}E}). In contrast, Mut2-CEA and wild-type Prot22-CEA showed comparable
overall RMSD, indicating that Y106F did not markedly alter the global protein stability
(\textbf{Fig.~S3A}). However, Mut2 exhibited reduced binding-pocket solvent-accessible surface
area (SASA), more persistent enzyme-substrate hydrogen-bond interactions, a shorter
catalytic distance, and a lower binding free energy (\textbf{Fig.~S3B--F}). These findings suggest
that Y106F enhances catalysis primarily by optimising the local pocket microenvironment
and stabilizing the substrate-binding conformation. Taken together, these results
demonstrate that the combined ProteinMPNN and ESM1b screening strategy can effectively
identify HF mutations that alter $\beta$-lactamase activity and substrate specificity.
Notably, mutations located in flexible regions appear more likely to represent sites
with elevated evolutionary potential.

\vspace{-3mm}
\subsection{Evolutionary Plasticity of High-Fitness Sites in $\beta$-Lactamases Across Contrasting Environmental Habitats}
\vspace{-2mm}

To investigate the evolutionary potential of naturally occurring
$\beta$-lactamases from distinct environmental sources, we applied the
high-fitness (HF) site screening strategy described above to identify
putative HF sites in representative sequences from each Ambler class. We
then assessed the variation at corresponding sites in natural sequences
through structural alignment. Wastewater and ocean habitats,
representing contrasting environmental conditions, were selected for
comparison. Sequences from each source were grouped by predicted Ambler
class, and one representative sequence for each class was selected by
structural clustering. HF sites were then predicted for each
representative sequence. The representative sequences were subsequently
structurally aligned against other sequences within the same structural
cluster, aligned residue pairs were extracted, and site entropy was
calculated for each aligned position.

\begin{figure}[!t]
  \centering
  \includegraphics[width=\linewidth]{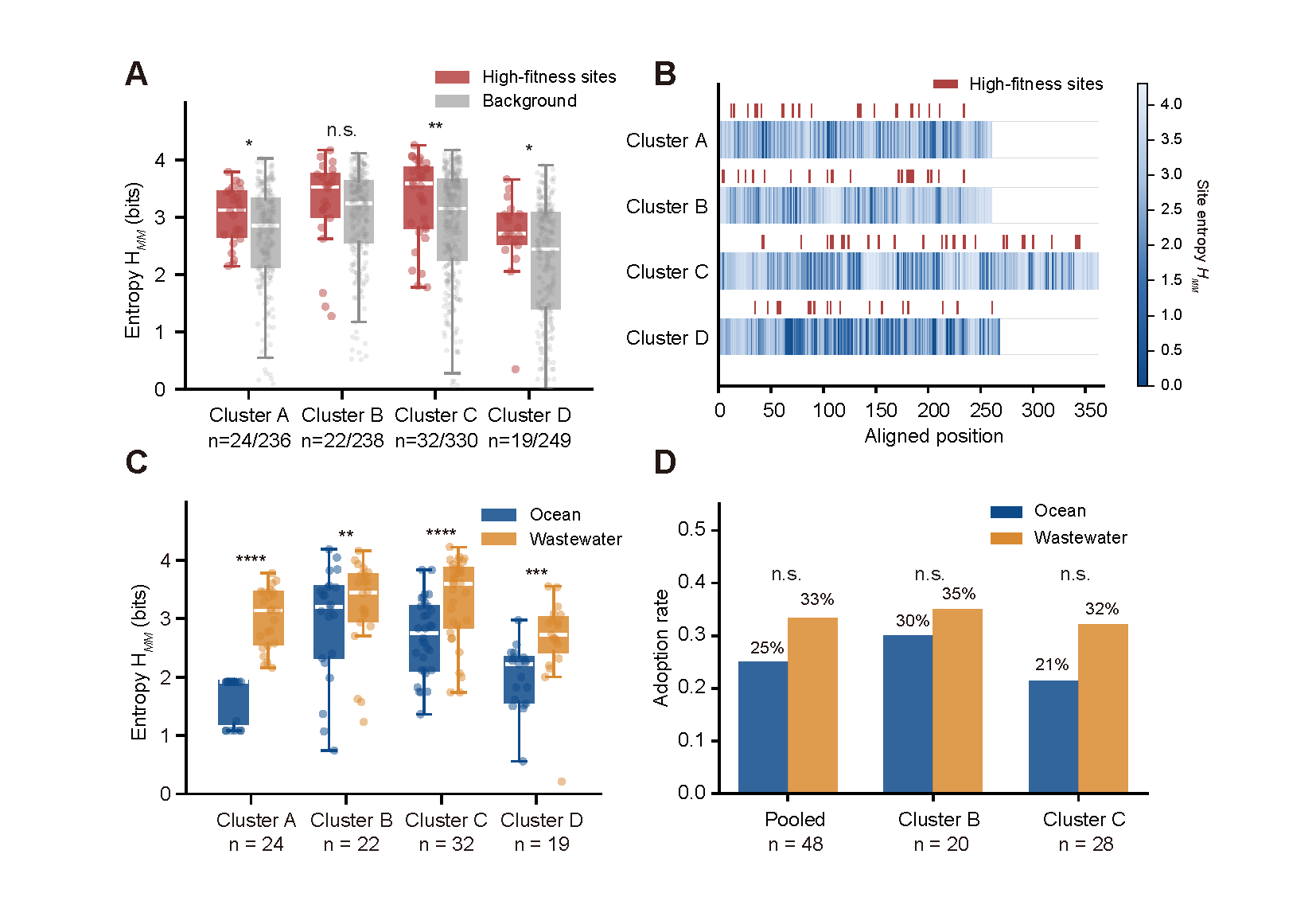}
    \setlength{\abovecaptionskip}{1pt}
    \caption{\scriptsize \textbf{Sequence variation at predicted high-fitness sites in
$\beta$-lactamases from contrasting habitats.} (A) Site entropy
($H_{\mathrm{MM}}$, bits) of predicted high-fitness
(HF) sites versus all non-predicted background positions for each
structure cluster. Values below clusters
give $n_{\mathrm{HF}}/n_{\mathrm{background}}$; differences were assessed with
one-sided Mann--Whitney U tests. \textbf{(B)} Locations of HF sites
along the site-entropy profiles of the representative sequences. Each
row corresponds to one representative sequence; every structurally
alignable position is coloured by its site entropy (dark blue,
conserved; pale, variable). \textbf{(C)} Comparison of entropy at
corresponding HF sites between wastewater- and ocean-derived sequences.
Differences were assessed with one-sided Wilcoxon signed-rank tests.
\textbf{(D)} Adoption rate, defined as the fraction of HF sites at which
the predicted HF amino acid is the dominant residue in that habitat.
Differences between habitats were assessed with two-sided McNemar exact
tests. In \textbf{C} and \textbf{D}, values below clusters
give $n_{\mathrm{HF}}$. $p$-values are represented in each plot as
$^{*}\mathit{p} < 0.05$, $^{**}\mathit{p} < 0.01$,
$^{***}\mathit{p} < 0.001$, $^{****}\mathit{p} < 0.0001$, and n.s.,
not significant ($\mathit{p} \geq 0.05$).}
  \label{fig:evolutionary_plasticity}
\end{figure}

Across the four representative sequences, a total of 97 HF sites were identified.
Compared with all non-predicted background positions, HF sites showed significantly
higher site entropy (median $H_{\mathrm{MM}} = 3.20$ vs.\ 2.93; $\mathit{p} = 1.2 \times 10^{-4}$;
$n = 97$ vs.\ 1,053). The same direction of effect was observed within each Ambler
class (median $H_{\mathrm{MM}}$: class A, 3.12 vs.\ 2.85; class B, 3.52 vs.\ 3.24;
class C, 3.59 vs.\ 3.15; class D, 2.71 vs.\ 2.44), reaching significance in classes
A ($\mathit{p} = 4.8 \times 10^{-2}$), C ($\mathit{p} = 3.8 \times 10^{-3}$), and D
($\mathit{p} = 3.6 \times 10^{-2}$), with a consistent but non-significant trend in class B
($\mathit{p} = 6.9 \times 10^{-2}$) (\textbf{Fig.~\ref{fig:evolutionary_plasticity}A}).
Cumulative distribution analysis further
revealed that the entropy distributions of HF and background sites differed
significantly ($D = 0.19$, $\mathit{p} = 1.2 \times 10^{-3}$), with HF sites shifted towards
higher entropy values overall (\textbf{Fig.~S4}). These findings indicate that the
enrichment of variation at HF sites is not driven by a minority of extreme
outliers, but is instead a broadly distributed phenomenon spanning multiple enzyme
classes. Site-entropy profiling showed that all 97 HF sites fell within highly
variable regions, with some sites occurring in contiguous clusters. This observation
suggests that these plastic regions may participate in the functional regulation
and adaptive evolution of $\beta$-lactamases as local sequence segments
(\textbf{Fig.~\ref{fig:evolutionary_plasticity}B}). Taken together, HF sites showed a greater tendency to be located in
regions tolerant of sequence variation rather than at highly conserved functional
residues, which reflects that this enzyme family achieves evolutionary adaptation
primarily through localized, non-disruptive functional modulation.

Comparison of site entropy across the four Ambler classes from the two
environmental sources revealed that wastewater-derived enzymes exhibited
significantly higher site entropy than ocean-derived enzymes in all four
classes (A: $\mathit{p} = 6.0 \times 10^{-8}$; B: $\mathit{p} = 4.2 \times 10^{-3}$;
C: $\mathit{p} = 3.5 \times 10^{-7}$; D: $\mathit{p} = 2.6 \times 10^{-4}$)
(\textbf{Fig.~\ref{fig:evolutionary_plasticity}C}). This pattern suggests that the more
complex and variable antibiotic and pollutant exposures characteristic
of wastewater environments may intensify localized adaptive selection,
resulting in higher amino acid substitution frequencies and greater
sequence plasticity at the relevant sites. By contrast, in the ocean
environment, the same sites appear to lack the evolutionary driving
forces that promote the diversification of positions.

To further assess the correspondence between predicted HF mutations and
natural sequence variation, we calculated the proportion of predicted HF
mutations that matched the dominant amino acid at the corresponding
position in sequences from the same environment. Because fewer than ten
class A and class D proteins were available in the ocean samples, this
analysis was restricted to 48 HF sites in class B and class C proteins,
for which both habitat groups had sufficient sample sizes. The
proportions of predicted HF mutations matching the dominant amino acid
at the corresponding positions were approximately 25\% in ocean-derived
sequences and 30\% in wastewater-derived sequences
(\textbf{Fig.~\ref{fig:evolutionary_plasticity}D}).
Although the proportion was consistently higher in wastewater-derived
sequences, the difference was not statistically significant in either
the pooled or class-specific analyses. Thus, within the current dataset,
we found no statistically significant evidence that habitat type
consistently biases the specific amino acid substitution favored at
individual HF sites. These results nevertheless suggest that
environmental conditions may be associated with differences in the
extent of sequence variation and plasticity at HF sites, highlighting
that environmental adaptation in $\beta$-lactamases may be reflected primarily
in the breadth of permissible sequence variation rather than in
convergence on a single substitutional trajectory.

\section{Discussion}

The growing availability of environmental metagenomic data has greatly
expanded the known diversity of microbial proteins, yet translating this
sequence diversity into functional and evolutionary understanding
remains a substantial challenge~\cite{Tas2021MetagenomicTools,Pavlopoulos2023DarkMatter}. In this study,
we developed a deep learning-guided framework that integrates PLMs,
ensemble learning, and structure-informed mutation prediction to
investigate environmental $\beta$-lactamases. By combining large-scale
sequence mining with experimental validation, we identified divergent
$\beta$-lactamase candidates capable of conferring resistance to multiple
$\beta$-lactam antibiotics and demonstrated that individual amino acid
substitutions can substantially alter their resistance phenotypes.
Further comparative analyses revealed differences in sequence
variability at predicted mutation-sensitive sites across environmental
habitats. These findings illustrate how AI-assisted approaches can
connect unexplored environmental protein diversity with experimentally
observable functional variation, providing great insights into the
potential adaptability of naturally occurring resistance enzymes.

One of the key challenges in exploring environmental antibiotic
resistance is identifying functional resistance enzymes that are highly
divergent from known sequences~\cite{Larsson2026ARGAnalyses}. Conventional
sequence-based approaches such as homology searches and profile-based
methods, have been instrumental in characterizing antibiotic resistance
genes~\cite{Bortolaia2020ResFinder}. However, their effectiveness can become
limited especially when candidate sequences diverge substantially from
experimentally characterized proteins~\cite{Schnoes2009AnnotationError,ArangoArgoty2018DeepARG}. PLMs
offer a complementary strategy by representing proteins in a learned
sequence space that captures information beyond direct pairwise
similarity~\cite{Rives2021Scaling}. In this work, the combination of
ESM2-derived representations with ensemble learning enabled the
systematic classification of candidate $\beta$-lactamases recovered from
environmental metagenomic datasets (\textbf{Fig.~\ref{fig:phylogenetic_tree}}). Importantly,
experimental characterization of two class C candidates, Prot22 and
Prot94, which shared only approximately 30\% sequence identity with
their closest characterized homologues, demonstrated their ability to
confer resistance to multiple $\beta$-lactam antibiotics in a heterologous
host (\textbf{Fig.~\ref{fig:experimental_validation}}). These findings support the value of combining
profile-based screening with learned protein representations to
prioritize divergent sequences for functional investigation. The
ecological distribution of the predicted candidates further revealed
substantial variation in $\beta$-lactamase class composition across
environmental habitats (\textbf{Figs.~\ref{fig:global_distribution} and
\ref{fig:habitat_distribution}}). Although these
patterns may partly reflect differences in sampling intensity, microbial
community composition, and database representation, they underscore the
importance of environmental microbial communities as sources of
$\beta$-lactamase diversity that remains incompletely characterized.

Beyond identifying existing resistance functions, an important question
is whether computational approaches can help explore the functional
consequences of mutations in previously uncharacterized enzymes. PLMs
and structure-conditioned sequence design provide complementary
perspectives on this problem: the former captures statistical
constraints associated with natural sequence variation, whereas the
latter identifies sequence alternatives compatible with a given protein
structure~\cite{Dauparas2022ProteinMPNN,Meier2021ZeroShot}. By integrating ESM1b with
ProteinMPNN, we identified candidate substitutions in Prot22 and Prot94
and experimentally evaluated their effects on antibiotic resistance.
Several single-amino-acid substitutions produced substantial changes in
resistance phenotypes, including enhanced resistance to particular
antibiotic classes and shifts in substrate-associated resistance
profiles (\textbf{Fig.~\ref{fig:mutation_validation}}). Notably, Y106F in Prot22 and S73T in
Prot94 were associated with pronounced phenotypic changes, suggesting
that considerable functional variation can arise from limited
modifications to otherwise conserved protein scaffolds. Molecular
dynamics simulations further suggested that these substitutions may
influence substrate positioning, local conformational stability, and
interactions within the catalytic pocket (\textbf{Fig.~\ref{fig:MD_analysis}}). These
structural interpretations remain provisional, as altered binding
interactions do not necessarily translate directly into increased
catalytic turnover. Nevertheless, the agreement between computationally
prioritized mutation sites and experimentally observed phenotypic
changes supports the utility of combining sequence- and structure-based
models to investigate the mutational potential of divergent
environmental enzymes. Importantly, the identification of mutations that
alter resistance does not by itself establish their evolutionary
accessibility or selective advantage under natural conditions, but it
provides experimentally testable hypotheses about the functional
flexibility of environmental $\beta$-lactamases.

The evolutionary significance of this mutational flexibility becomes
particularly interesting when considered in an ecological context.
Environmental $\beta$-lactamases occur across habitats that differ in
microbial community structure, chemical exposure, and potential
selective pressures~\cite{Deak2026SystematicReview}. Our comparative analysis showed
that computationally predicted high-fitness sites exhibited greater
sequence variability than background positions, suggesting that these
sites tend to occur in regions with relatively high tolerance for amino
acid variation. Furthermore, the corresponding positions in
wastewater-derived sequences displayed higher site entropy than those in
ocean-derived sequences across the four Ambler classes
(\textbf{Fig.~\ref{fig:evolutionary_plasticity}}). One possible explanation is that the heterogeneous chemical and
microbial conditions associated with wastewater environments favour the
maintenance of greater sequence diversity~\cite{Ju2019Resistomes}. However,
higher sequence entropy cannot be interpreted as direct evidence of
positive selection, and differences in phylogenetic composition,
sampling depth, or environmental representation may also contribute to
the observed patterns. Moreover, the absence of significant
habitat-specific differences in the correspondence between predicted
substitutions and dominant naturally occurring residues suggests that
greater sequence variability does not necessarily imply convergence
towards particular amino acid states. These observations therefore
support an association between environmental origin and sequence
variability at computationally identified mutation-sensitive positions,
while leaving open the question of whether and how local environmental
pressures shape the evolutionary trajectories of these enzymes.

Several limitations should be acknowledged. First, the generalizability
of the ensemble-learning framework to highly divergent proteins require
further validation. Second, experimental characterization was limited to
two class C candidates, with resistance assessed primarily by MIC
measurements; broader functional assays across Ambler classes are needed
to distinguish catalytic activity from effects of protein expression and
stability. Third, mutation predictions were evaluated using a limited
number of single substitutions, whose effects may depend on sequence
background and antibiotic selection. Finally, ecological comparisons
relied on available metagenomic sequences rather than direct
measurements of selection pressures or longitudinal evolutionary
changes. Future studies integrating broader experimental validation,
experimental evolution, and environmentally resolved metagenomic data
will help clarify how predicted mutational potential translates into
realized evolutionary adaptation. Despite these limitations, our study
establishes an integrated framework combining protein language models,
structure-informed sequence exploration, and experimental
characterization to investigate the functional diversity of
environmental $\beta$-lactamases. More broadly, this framework provides a
foundation for exploring how uncharacterized environmental enzymes may
contribute to the emergence and diversification of antibiotic
resistance.

\section*{Funding}

This work was supported by the National Natural Science Foundation of China
(32170079, 32200035), the Natural Science Foundation of Guangdong Province
(2024A1515012593 and 2026A1515011616), the Guangdong S\&T Program
(2024B1111130001), and the Guangdong Talent Scheme (2021QN020100).

\section*{Conflicts of Interest}

The authors declare no conflicts of interest.

\section*{Data Availability Statement}

The data that support the findings of this study are available upon
reasonable request.


\bibliographystyle{splncs04}
\bibliography{references}

\end{document}